\documentclass[aps,prl,reprint]{revtex4-1}
\usepackage{graphicx}
\usepackage{bm}
\usepackage{braket}
\usepackage{amsmath}
\usepackage{mathrsfs}
\usepackage{float}
\usepackage{color}
\usepackage{hhline}
\usepackage{tabularx} 
\newcolumntype{Z}{>{\centering\let\newline\\\arraybackslash\hspace{0pt}}X}
\renewcommand{\thispagestyle}[1]{}

\begin{document}
\title{The Role of Tensorial Electronic Friction in Energy Transfer at Metal Surfaces}
\author{Mikhail Askerka}\thanks{M.A. and R.J.M. made equal contribution to this work}
\author{Reinhard J. Maurer}\thanks{M.A. and R.J.M. made equal contribution to this work}
\author{Victor S. Batista}
\affiliation{Department of Chemistry, Yale University, New Haven CT 06520, United States} 
\author{John C. Tully}
\email[]{john.tully@yale.edu}
\affiliation{Department of Chemistry, Yale University, New Haven CT 06520, United States}

\date{\today}

\begin{abstract}
An accurate description of nonadiabatic energy relaxation is crucial for modeling atomistic dynamics at metal surfaces. Interfacial energy transfer due to electron-hole pair excitations coupled to motion of molecular adsorbates is often simulated by Langevin molecular dynamics with electronic friction. Here, we present calculations of the full electronic friction tensor by using first order time-dependent perturbation theory (TDPT) at the density functional theory (DFT) level. We show that the friction tensor is generally anisotropic and non-diagonal, as found for hydrogen atom on Pd(100) and CO on a Cu(100) surfaces. This implies that electron-hole pair induced nonadiabatic coupling at metal surfaces leads to friction-induced mode coupling, therefore opening an additional channel for energy redistribution. We demonstrate the robustness and accuracy of our results by direct comparison to established methods and experimental data.
\end{abstract}

\maketitle 
The rates and pathways of energy flow can be controlling factors in dynamical processes at surfaces, such as adsorption, desorption, surface diffusion, and chemical reaction~\cite{Arnolds2011}. In particular, at metal surfaces the excitation and de-excitation of electron-hole pairs (EHP) can be a contributing or even dominant mechanism for energy transfer. Electronic transitions driven by nuclear motion represent a violation of the adiabatic (Born-Oppenheimer) principle. Nevertheless, nonadiabatic transitions are common in metals due to the presence of a continuum of low-lying electronic states~\cite{Wodtke_Tully_2004}.  Nonadiabatic phenomena have been observed experimentally in  molecular-beam surface scattering~\cite{Rettner_Auerbach_PRL_1985,Cooper_Wodtke_Chem_Sci_2010,Watts_Sitz_Surf_Sci_1997}, inelastic electronic tunneling spectroscopy~\cite{Sainoo_Kawai_PRL_2005}, and sum frequency generation spectroscopy~\cite{Bartels_Wodtke_2011,Arnolds_Bonn_Surf_Sci_Rep_2010}. Atomistic simulations of these processes are challenging since they require methods that go beyond adiabatic molecular dynamics. In this context, rigorous quantum dynamics or mixed quantum-classical Ehrenfest~\cite{Li2005,Horsfield_Fisher_2006} or surface hopping dynamics~\cite{Tully1990,Tully_JCP_2012} methods can be used, but they are computationally expensive.

Molecular dynamics with electronic friction (MDEF) is an approach that has been widely used to study EHP-induced adsorbate/surface energy transfer in dynamics at surfaces~\cite{Headgordon_Tully_JCP_1995,Kindt_Tully_JCP_1998}. In the MDEF approach, atomic positions evolve according to classical mechanics Langevin equation, with forces obtained for a potential energy surface, $V\mathbf{(R)}$, and energy dissipation due to EHP excitations incorporated with a friction $\mathbf{\Lambda}$:
\begin{align}\label{eq-langevin}
M\ddot{R_i}= - \frac{\partial{V\bm{(R)}}}{\partial{R_i}} - \sum_{j} \Lambda_{ij}\dot{R_j} + \mathscr{R}_i(t).
\end{align}
$\Lambda_{ij}$ is an element of the (3N$\times$3N)-dimensional electronic friction tensor $\mathbf{\Lambda}$ where N is the total number of atoms that are explicitly considered in the dynamics simulation of the combined adsorbate/surface system. Here, we, however, focus on EHP coupling effects induced by the adsorbate motion and therefore only consider the adsorbate atoms explicitly. $\mathscr{R_i}(t)$ is a random fluctuating force determined by an effective 'electronic friction' \emph{via} the fluctuation-dissipation theorem that ensures detailed balance~\cite{Headgordon_Tully_JCP_1995,kubo2012statistical}. The second term in the r.h.s. of eq. \ref{eq-langevin} describes the friction forces acting on the coordinate $i$ due to motion in $j$ direction. In the Langevin picture, nonadiabatic electronic friction introduces hydrodynamic coupling effects on the motion of molecular adsorbates similar to those in diffusion~\cite{Kraft_PRE_2013,Reichert_PRE_2004}.
The electronic friction constants $\Lambda_{ij}$ can be obtained from electronic structure calculations, usually based on Density Functional Theory (DFT)~\cite{Hellsing1984,Headgordon_Tully_JCP_1992,Bauer1998} or Many-Body Perturbation Theory~\cite{Faber2011}. Another commonly used approach is the so-called Local Density Friction Approximation (LDFA)~\cite{Juaristi_PRL_2008,Li_Yinggang_PRL_1992, Blanco-Rey_Juaristi_PRL_2014}, which has been derived for a single ion experiencing the stopping power of a homogeneous free electron gas~\cite{Echenique_Nieminen_SSC_1981,Echenique_Nieminen_PRA_1986}. Typically, all of the above methods include an incomplete account of electronic friction based on diagonal or isotropic tensors, calculated either along Cartesian crystallographic directions or normal mode displacement coordinates, therefore neglecting couplings between different degrees of freedom.

In the present work, we combine first order time-dependent perturbation theory (TDPT) and DFT to assess the importance of the tensorial properties of electronic friction. We focus on hydrogen atom motion on a Pd(100) surface and carbon monoxide (CO) vibrational cooling on a Cu(100) surface. We find that the off diagonal elements of the friction tensor can be comparable in magnitude to the diagonal elements. Furthermore, we find that the friction tensor and Hessian do not share the same eigenvectors. Therefore, electronic friction in general couples the adsorbate vibrational modes and promotes intramolecular vibrational energy redistribution due to EHP excitations. 

We separate vibrational and electronic degrees of freedom, and assume nuclear harmonic wavefunctions, according to the standard methodology based on Fermi's Golden Rule. Nonadiabatic couplings define the friction tensor elements in the Kohn-Sham (KS) ground state orbitals as follows:
\begin{align}\label{eq-friction-tensor}
  \Lambda_{ij} &= \pi\hbar^2 \omega \sum_{\mathbf{k},\nu,\nu'} \braket{\psi_{\mathbf{k}\nu}|\mathbf{\nabla}_{i}\psi_{\mathbf{k}\nu'}}\cdot\braket{\psi_{\mathbf{k}\nu'}|\mathbf{\nabla}_{j}\psi_{\mathbf{k}\nu}}  \cdot \\ \nonumber &\cdot\delta(\epsilon_{\mathbf{k}\nu}-\epsilon_{\mathbf{k}\nu'}-\hbar \omega) .
\end{align}
where $\ket{\psi_{\mathbf{k}\nu}}$ and $\ket{\psi_{\mathbf{k}\nu'}}$ are KS states and the sum over $\mathbf{k}$ indicates Brillouin zone sampling~\cite{Headgordon_Tully_JCP_1992,Hellsing1984,Luntz_Makkonen_PRL_2009, supplemental}. KS eigenenergies are indicated as $\epsilon_{\mathbf{k}\nu}$. Derivatives with respect to nuclear Cartesian coordinates $i$ and $j$ are computed by finite differences. In order to facilitate convergence of the spectral density we approximate the delta function with a normalized Gaussian distribution of finite width $\sigma$ centered at the Fermi level~\cite{Methfessel1989} (see Supporting Information (SI)). We express $\braket{\psi_{\mathbf{k}\nu'}|\mathbf{\nabla}_{\mathbf{R}}\psi_{\mathbf{k}\nu}}$ through the generalized eigenvalue relation as derivatives of the Hamiltonian and overlap matrices in a local atomic orbital basis  representation~\cite{Headgordon_Tully_JCP_1992,Savrasov1996} (\emph{cf.} SI). The tensorial property of $\mathbf{\Lambda}$ as introduced by eq. \ref{eq-langevin} accounts for energy transfer in coordinate $i$ due to EHP excitations induced by the motion along coordinate $j$.

\begin{figure}[t]
\includegraphics[width=\linewidth]{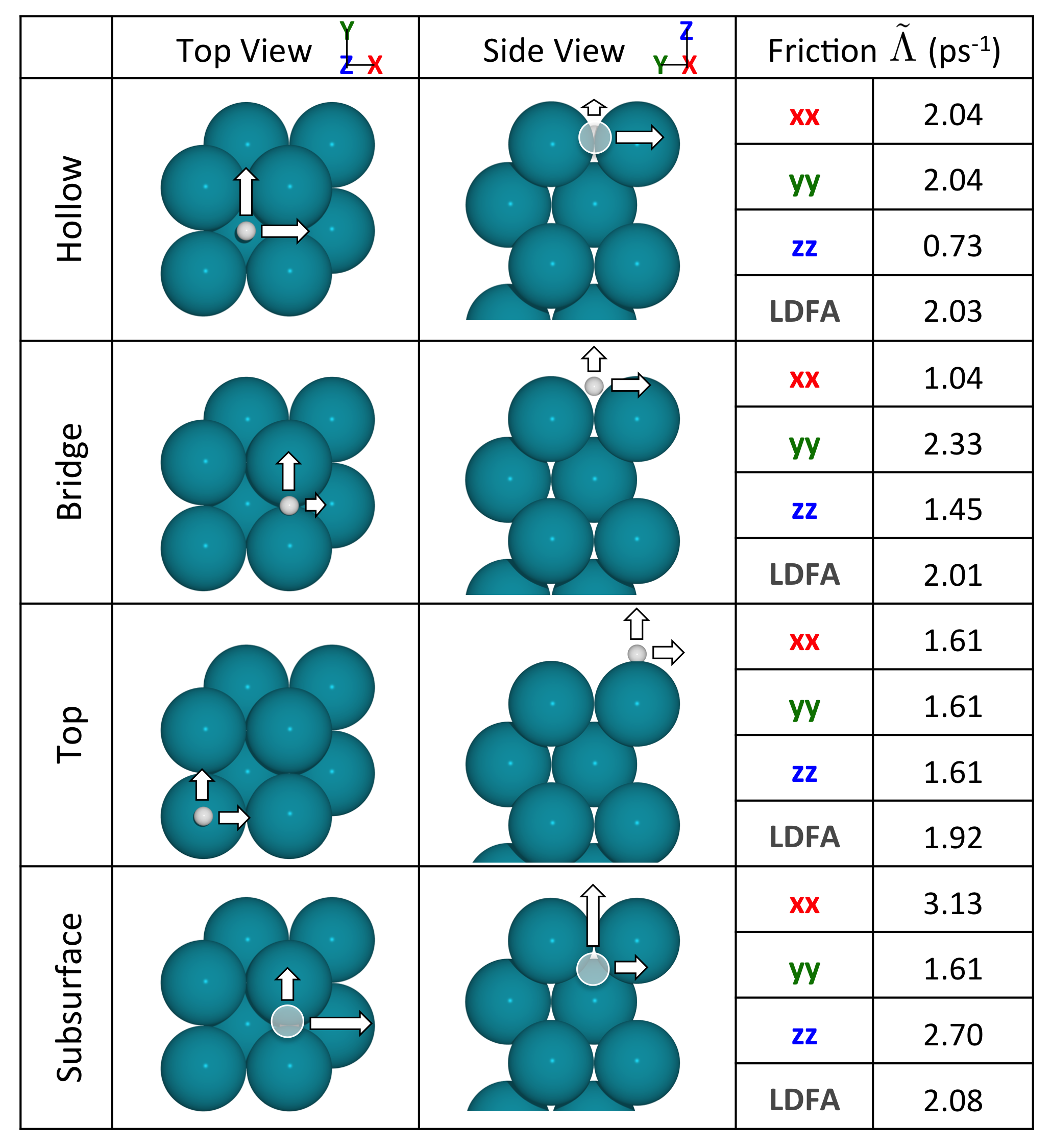}
\caption{(left) Hydrogen atom on Pd(100) as viewed from xy (top view) and yz (side view) planes for the hollow, bridge, top, and subsurface sites. Dimmed circles point at the positions of the Hydrogen atom when it is not directly visible at current view. The depicted arrows are proportional to the magnitude of electronic friction. (right) Components of the mass-weighted friction tensor (in ps$^{-1}$) along Cartesian directions and the isotropic rate as given by LDFA $\tilde{\Lambda}_\mathrm{LDFA}$.}
\label{fig:H_sites}
\end{figure}

Assuming exponential decay, we can calculate the EHP-induced relaxation rate~\cite{Headgordon_Tully_JCP_1992,Head-Gordon1992} $\Gamma_{\omega}$ of a given vibrational frequency as follows~\cite{Butler1979,Hellsing1984}:
\begin{align}\label{eq-lifetime}
 \Gamma_{\omega} =\mathbf{e}^T_{\omega}\cdot\mathbf{\tilde{\Lambda}}\cdot\mathbf{e}_{\omega},
\end{align}
where $\mathbf{\tilde{\Lambda}}$ refers to the mass-weighted Cartesian friction tensor (relaxation rate tensor)  $\tilde{\Lambda}_{ij}=\Lambda_{ij}/(\sqrt{m_i}\sqrt{m_j})$ and $\mathbf{e}_{j}$ refers to the displacement vector of the vibrational motion in question. The vibrational lifetime is the inverse of the corresponding relaxation rate $\tau_{\omega} =  1/\Gamma_{\omega}$ at a given geometry. 

In the following we analyze the structure and elements of the Cartesian electronic friction tensor using two model systems: A hydrogen atom adsorbed on a Pd(100) surface and CO adsorbed on a Cu(100) surface. We use a semi-local DFT formulated by Perdew, Burke, and Ernzerhof (PBE)~\cite{PBE}, as implemented in the periodic local atomic-orbital package SIESTA~\cite{Soler2002} (convergence with respect to all computational settings has been validated~\cite{settings}). We compare our results with calculations based on the LDFA method, which by construction neglects directionality and hydrodynamic coupling in its description of electronic friction.

Fig. \ref{fig:H_sites} shows the components of the friction tensor along Cartesian coordinates (in this case coinciding with the crystallographic axes) for a H atom adsorbed at various sites of the Pd(100) surface. We considered adsorption at a hollow site, at non-equilibrium bridge and top sites, and at the tetrahedral subsurface site. Due to the symmetry of the adsorption sites the electronic friction tensor is a diagonal matrix, however it is generally anisotropic. Differences in the relaxation rates along different crystallographic directions can be significant. For instance, for the hollow adsorption site, friction along the x (y) direction is almost three times stronger than along the z direction. We find that substrate symmetry is correctly reflected in the friction tensor. For example, motion along the x and y directions are equivalent for hydrogen adsorbed on the hollow and atop sites with 4-fold symmetry, resulting in identical relaxation rates in the corresponding directions. In contrast, the bridge site with 2-fold symmetry exhibits different rates in all three Cartesian directions.

For a comparison, we calculated the relaxation rates according to the LDFA approximation, as given by the clean surface electron density at the binding site of the H atom and using tabulated phase shifts from Ref. \onlinecite{Puska_Nieminen_PRB_1983} (Figure  \ref{fig:H_sites})~\cite{supplemental}. By construction, LDFA yields a scalar, isotropic value of electronic friction which we can translate to almost a constant rate of about 5.6 to 6.3 ps$^{-1}$ for Hydrogen on the surface sites and 7.1 ps$^{-1}$ at the subsurface site. These rates are significantly larger than TDPT rates along the individual Cartesian directions, suggesting that electronic friction in LDFA may be systematically overestimated when compared to TDPT.
\begin{figure}[t]
\includegraphics[width=\linewidth]{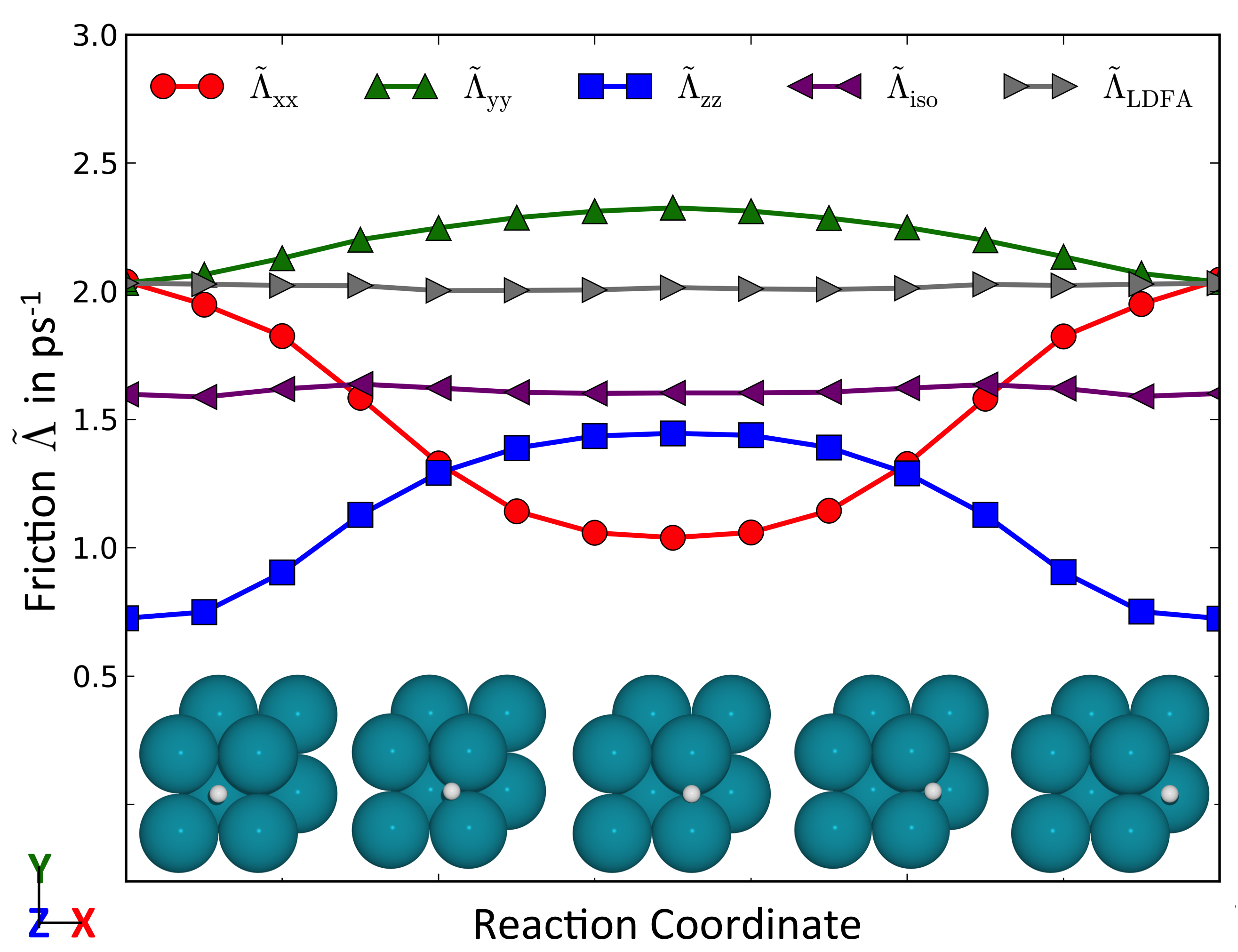}
\caption{Cartesian components of the mass-weighted friction tensor $\mathbf{\tilde{\Lambda}}$ (in ps$^{-1}$) of a Hydrogen atom on Pd(100) as it moves from one equilibrium hollow site to another (first and last points) across a bridge site (middle point). Shown are the relaxation rates along the three Cartesian components, the rate as given by the average of trace of the mass-weighted friction tensor ($\tilde{\Lambda}_{\mathrm{iso}}=\mathrm{Tr}(\mathbf{\tilde{\Lambda}})/3$), and the isotropic rate as given by LDFA $\tilde{\Lambda}_\mathrm{LDFA}$.}
\label{fig:H_scan}
\end{figure}

To further explore the anisotropic effects of friction on dissipative dynamics, we calculated the Cartesian components of the friction tensor of H bound to Pd(100) as it is displaced along the minimal energy path for diffusion from one equilibrium hollow site to another, across a bridge site (Figure \ref{fig:H_scan}). Our TDPT results show that as the hydrogen atom moves from one hollow site to the other along the $x$ direction, the $zz$ and $yy$ components of the friction tensor grow and reach a maximum at the bridge site, whereas the $xx$ component simultaneously decreases and reaches a minimum at the bridge site. The $yy$ component is the least sensitive to the position along this minimum energy path; moreover, since the motion occurs in the $xz$ plane, and the $y$ component is decoupled from $x$ and $z$ components in the tensor (see SI, Figure S1), it would have a zero contribution to the energy loss along this path. It is important to note here that the friction tensor at positions in-between the high-symmetry sites does, however, show sizeable non-zero off-diagonal coupling between $x$ and $z$ components (about 30\% of the corresponding diagonal $xx$ component) along this trajectory (see SI, section II for details). 

In contrast, electronic friction given by LDFA  remains essentially constant along the path. This is expected, since, according to the LDFA method, the electronic friction is simply a function of the local electron density created by the Pd surface, which is almost constant along the minimum energy path. Interestingly, the average of the TDPT dissipation rates along different directions with $\tilde{\Lambda}_{\mathrm{iso}}$ is also largely constant along the path, although it is considerably smaller than the rate given by LDFA. This could point to a fundamental connection between the trace of the friction tensor and the electron density, however, this needs to be further investigated. Direct comparison between tensorial friction based on TDPT and LDFA could motivate future approximate methods that go beyond LDFA by, for example, including the correlation of electronic friction on the local density and its gradient.

We note that the dynamics along the minimum energy path will follow a velocity vector with non-zero components along $x$ and $z$ directions. Therefore, the energy loss at or close to the bridge site will be slower than around the hollow sites. Additionally, the non-zero off-diagonal elements in TDPT along this diffusion path will contribute to an unequal change of velocity adding to velocity changes that are already imposed by the PES corrugation. None of the above effects are captured by the simplified description of $\mathbf{\tilde{\Lambda}}$ given by LDFA. From these data, however, it is difficult to conclude whether the off-diagonal couplings significantly alter the outcome of an individual adatom trajectory or if they can be safely neglected.

\begin{figure}[t]
\includegraphics[width=\linewidth]{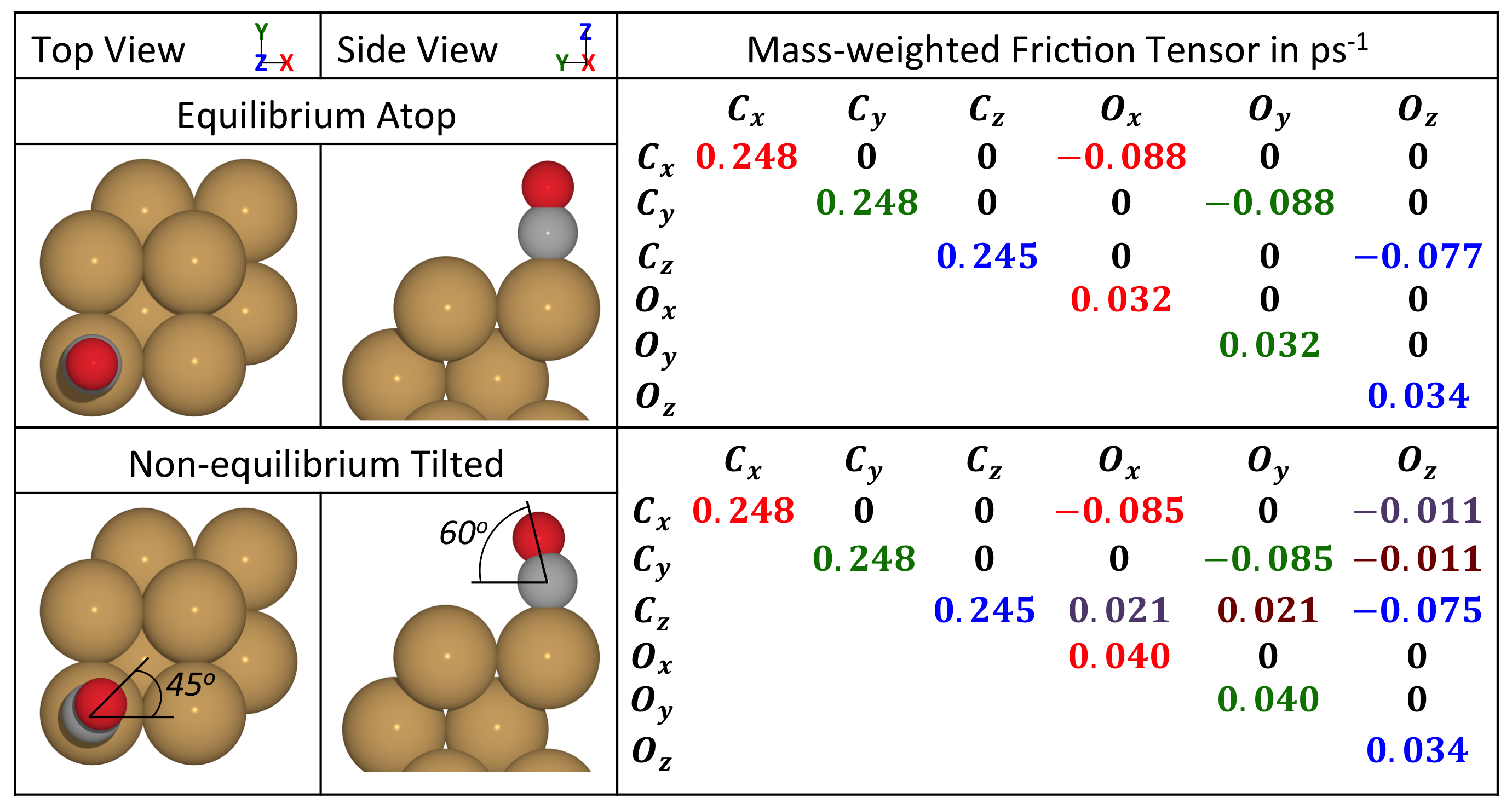}
\caption{(left) CO adsorbed at an atop site of a Cu(100) surface in the equilibrium position and a non-equilibrium tilted geometry as viewed from xy (top view) and yz (side view) planes. (right) Mass-weighted friction tensor (in ps$^{-1}$) for the two geometries. The coloring scheme is consistent for x (red), y (green), z (blue), xz (purple), and yz (brown) components of the friction tensor. Components smaller than 0.002 ps$^{-1}$ are set to zero. }
\label{fig:CO_tensor}
\end{figure}

The relevance of off-diagonal elements in the friction tensor becomes much more evident when analyzing molecular adsorbates on metal surfaces. 
We reconsider the well-studied model system of CO adsorbed on the Cu(100) surface~\cite{Morin_1992,Head-Gordon1992,Krishna_Tully_2006} by calculating the full (6$\times$6) Cartesian friction tensor for CO adsorbed in the equilibrium upright position and also in a non-equilibrium tilted position at the atop site (see Figure \ref{fig:CO_tensor}). We find that even at a high symmetry site, the friction tensor for the CO molecule has significant off-diagonal components. In the upright position, Oxygen and Carbon atom components only couple within the same Cartesian direction: $C_x$ is coupled to $O_x$ and so on. The magnitude of off-diagonal elements is on the same level as the magnitude of diagonal elements (Figure \ref{fig:CO_tensor}). We also find that these elements can be negative (as long as the overall matrix remains positive-definite), effectively describing a reduction of electronic friction as a result of non-vanishing velocity in the coupled directions. For the general case of non-equilibrium sites such as the here presented tilted CO position, the friction tensor may be fully populated.

Calculating the vibrational lifetime along normal mode vibrations ('Along NM' in Table I) for CO on Cu(100) using eq. \ref{eq-lifetime} we find normal mode lifetimes in good agreement with previous TDPT studies of our group using lifetimes based on cluster models~\cite{Head-Gordon1992} and periodic plane-wave calculations~\cite{Krishna_Tully_2006}. In addition, we find a fair agreement with the experimental reference for the lifetime of the internal stretch (IS) of 2$\pm$1~ps measured by Morin \emph{et al}~\cite{Morin_1992}. We also simulated the explicit dynamical energy loss along the IS mode  by integrating the Langevin equation along the corresponding one-dimensional PES (see SI section III for details). The resulting vibrational lifetime of 3.92~ps is almost identical to the static instantaneous lifetime of 4.07~ps. We also find that another short lived vibrational mode is the frustrated rotation (3.61~ps), which is in accord with previous theoretical and experimental work~\cite{Hirschmugl1990109}. 

Our calculations only consider vibrational energy loss due to EHP excitations, we therefore expect them to yield an upper limit to the experimentally observed vibrational lifetime, notwithstanding all other approximations taken. LDFA results for the IS lifetime in literature lie above or below our result, depending on the construction of the embedding density~\cite{Rittmeyer_Meyer_PRL_2015,Tremblay_Monturet_PRB_2010}. The applicability of LDFA for molecular adsorbates has been a topic of recent discussion~\cite{Juaristi_PRL_2008, Luntz_Makkonen_PRL_2009, Rittmeyer_Meyer_PRL_2015}. Table I shows results for an embedding density based on the density of the clean surface (Independent Atom Approximation, LDFA-IAA) and an embedding density as given by the atoms-in-molecules scheme (LDFA-AIM)~\cite{Rittmeyer_Meyer_PRL_2015,Novko2015}. Analogously to the analysis of hydrogen, we can calculate atom-wise isotropic friction coefficients from the TDPT tensor by averaging over diagonal elements and discarding off-diagonals. The resulting IS lifetime of 5.8~ps is in close agreement with the LDFA-IAA value of 6.1~ps.

\begin{table}
\caption{Vibrational lifetimes (in ps) of the four principle displacement modes for CO adsorbed at the top site of a  Cu(100) surface as calculated by eq. \ref{eq-lifetime}. 'Along NM' refers to lifetimes calculated along the vibrational normal modes. 'Along FM' refers to lifetimes calculated along the eigenvectors of the mass-weighted friction tensor. IS: internal stretch mode, SA: surface-adsorbate stretch mode, FT: frustrated translation mode, FR: frustrated rotation mode.}\label{tab:table_comparison}
\begin{tabularx}{\linewidth}{ZZZZZ} \hhline{=====} \noalign{\vskip 0.1cm}   
      & IS & SA & FT & FR \\ \hline  \noalign{\vskip 0.1cm}   
      Along NM & 4.07(3.92)$^{a}$ & 17.0 & 70.5& 3.61\\
      Along FM & 3.70              & 108  & 808 & 3.58\\  \hline  \noalign{\vskip 0.1cm}   
      Ref \onlinecite{Krishna_Tully_2006}  &3.3 &13.7 &19.5 &3.8 \\ 
      Ref \onlinecite{Headgordon_Tully_JCP_1992}  & 3.3& 82& 108&2.3 \\ 
      Ref \onlinecite{Rittmeyer_Meyer_PRL_2015}  & 6.1 & \multicolumn{3}{c}{LDFA-IAA}  \\       
      Ref \onlinecite{Rittmeyer_Meyer_PRL_2015}  & 1.8 & \multicolumn{3}{c}{LDFA-AIM}  \\ 
      Exp. & 2 $\pm$ 1 \cite{Morin_1992}&  & & $\geq$ 1 \cite{Hirschmugl1990109}\\[0.01cm] \hhline{=====}
\end{tabularx}
$^a$: Lifetime in parentheses was calculated following the explicit Langevin dynamics along the internal stretch mode. For details see SI, Section III.
\end{table}

For the case of CO on Cu(100), we find that EHP-induced vibrational relaxation is not only mode-selective, but also introduces vibrational mode coupling. Whereas adsorbate motion along normal modes has often been assumed to transform the friction tensor into a diagonal form~\cite{Kindt_Tully_JCP_1998}, by spectral analysis of the mass-weighted friction tensor for CO adsorbed atop on Cu(100) (Fig \ref{fig:CO_tensor}) we find eigenvectors that are not identical to the vibrational normal modes (See web enhanced object in the SI for an interactive animation~\cite{web_enh}). Unlike vibrational normal modes, eigenmodes of the friction tensor ('friction modes') are not required to preserve the center of mass. Characterizing friction modes of upright CO on Cu(100) by projection with vibrational normal modes, we find that frustrated rotation (FR) and frustrated translation (FT) modes only minimally deviate from their normal mode analogues, however the internal stretch mode (IS) and the surface-adsorbate (SA) stretch mode are strongly mixed. This becomes clear from considering the differences in vibrational lifetime along normal modes and along friction modes as calculated according to eq. \ref{eq-lifetime}. The lifetime of the friction mode closest to the IS is slightly reduced compared to the internal stretch, whereas the surface-adsorbate (SA) stretch lifetime is an order of magnitude larger. The consequence of this effective mode coupling is that vibrational excitation of individual adsorbate normal modes will induce EHP-mediated energy-loss that in turn facilitates vibrational mode coupling between adsorbate degrees of freedom, effectively introducing faster energy redistribution. Such EHP-mediated vibrational mode coupling may be a contributing factor in electron-tunneling induced adsorbate migration as observed for CO on Pd(110)~\cite{Komeda2002}.

In summary, we have shown how to account for anisotropic electronic effects in the description of energy relaxation of molecular adsorbates at metal surfaces. Our calculations, combining first-order perturbation theory and ground state DFT for H on Pd(100) and CO on Cu(100), demonstrate that the electronic friction tensor of adatoms and molecular adsorbates is (1) anisotropic and vibrational motions are damped differently in different directions and (2) that it is fully populated and generally not diagonal in either normal mode or crystallographic coordinates. In comparison to LDFA we find significant deviations. Our findings show that electronic friction leads to enhanced intramolecular vibrational energy redistribution among otherwise weakly coupled vibrational modes of the molecular adsorbates. Our methodology is completely general and could be extended beyond ground-state DFT~\cite{Abad_JCP_2013}. When coupled with an analytical description of non-adiabatic coupling matrix elements~\cite{Baroni2001,Savrasov1996}, it could be efficiently incorporated into molecular dynamics simulations.

\begin{acknowledgments}
JCT acknowledges financial support by the US Department of Energy - Basic Energy Science grant DE-FG02-05ER15677. VSB acknowledges financial support from the Air Force Office of Scientific research grant FA9550-13-1-0020 and high performance computing time from NERSC.
\end{acknowledgments}


\begin{thebibliography}{47}
\bibitem [{web()}]{web_enh}%
  \BibitemOpen
  \href@noop {} {}\bibinfo {note} {A web-enhanced interactive visualization of
  mode displacements can be accessed at prl.aps.org}\BibitemShut {NoStop}%
\end{thebibliography}
\end{document}